\documentclass{osa-article}

\journal{osac}

\articletype{Research Article}

\usepackage{lineno}

\begin{document}

\title{A Local Area Quantum Teleportation Network Based on an Array of Electrically Activated Graphene Waveguides}

\author{Muhammad Asjad,\authormark{1,2} Montasir Qasymeh,\authormark{1,*} and Hichem Eleuch\authormark{3,4}}

\address{\authormark{1}Electrical and Computer Engineering Department, Abu Dhabi University, Abu Dhabi 59911, UAE\\
\authormark{2}Department of Mathematics, Khalifa University, Abu Dhabi 127788, UAE\\
\authormark{3}Department of Applied Physics and Astronomy, University of Sharjah, Sharjah 27272, UAE\\
\authormark{4}Institute for Quantum Science and Engineering, Texas AM University, College Station, TX 77843, USA}

\email{\authormark{*}montasir.qasymeh@adu.ac.ae} 



\begin{abstract}
We present a scheme to generate a continuous variable (CV) multipartite entangled state using an array of plasmonic graphene waveguides that are activated by nonclassical driving microwave modes. Within this scheme, we can exploit the interaction of two light fields coupled to the same microwave mode in each waveguide to produce any type of multipartite Gaussian entangled state. A teleportation network is illustrated using the resultant CV multipartite entangled state. In particular, the proposed setup  enables coherent state teleportation across remotely connected nodes with fidelity above a threshold limit of 2/3, providing secure quantum teleportation networking even in the presence of losses. 
\end{abstract}

\section{Introduction}
High-fidelity transfer of quantum states between distant nodes is key to realizing quantum networks  \cite{transfer, internet,internet2}. However, the transfer fidelity is severely degraded by separating distances \cite{communication}. Despite the tremendous progress that has been made in developing quantum repeaters and relaying quantum information over long distances  \cite{repeater,repeater2}, quantum repeaters with practical modalities have yet to be implemented \cite{repeaterlimit,diamond}. However, small-scale quantum networks have been realized by implementing entanglements between distant quantum nodes. For instance, short-scale quantum state transfer with moderate separation distances (a few meters) has been reported using different platforms \cite{herald,heraldloss}. These schemes include quantum state transfer between quantum dots separated by 5 m \cite{spins}, trapped atoms separated by 20 m \cite{atoms}, and solid-state qubits separated by 3 m \cite{solid}. Also, quantum state transmission between distant microwave photons has been proposed using hybrid electro-optic entanglement \cite{EO}. Generating quantum entanglement using a graphene-based structure has been reported by our group a couple of years ago \cite{qasUS.2,qasIeee}. The proposed approach encompasses a superconducting capacitor that incorporates a plasmonic graphene waveguide. A microwave signal drives the capacitor while an intensive surface plasmon polariton (i.e., SPP) mode is launched to the graphene layer. The microwave voltage and the SPP mode interact by electrically modifying the graphene optical conductivity. It then follows that optical SPP sidebands are generated. This interaction is explained by noting that the graphene chemical potential is a nonlinear function of the driving microwave signal. Thus, by expressing the chemical potential by its series expansion, and considering the case of small quantum driving voltages, the chemical potential can be approximated up to the first order. Same approximation is implemented to the graphene conductivity and the effective permittivity \cite{qasUS1, qsasexpress}. Consequently, by using the approximated effective permittivity in the governing Hamiltonian, two entangled optical SPP sidebands are produced\cite{qasIeee}. Furthermore, we have demonstrated quantum optical state transfer between two separated nodes using the same graphene-loaded capacitor structure \cite{asjad21}. High-fidelity teleportation with lengths of up to 7 km was achieved using a free-space classical channel. Extending this approach to realize a quantum optical network has the potential to accomplish high-fidelity transmission among numerous nodes separated by substantial distances.  

In this study, we propose the use of an array of hybrid plasmonic graphene waveguides to generate a continuous variable (CV) multipartite Gaussian Greenberger-Horne-Zeilinger (GHZ)-like entangled state between remotely connected nodes. We illustrate how the generated CV multipartite entangled states can be exploited to provide teleportation between connected plasmonic graphene waveguides. The proposed configuration incorporates $N$-entangled beams through $N$-plasmonic waveguides, which are coupled via a suitable sequence of $N-1$ beam splitters (BSs). This result reveals that state transfer can be accomplished by performing multipartite Bell measurements on the N entangled beams. Our calculations demonstrate that a quantum state transfer can be realized within a network of up to 3 (4) nodes connected in a star configuration with 20 (2.3)-km separation distances by using reasonable specifications of a free-space channel (with an attenuation of 0.005 dB/Km). We note that proposals considering two nodes with significant teleportation distances (tens of kms) have been reported. However, in the current study, we are investigating the scheme of having N remotely connected nodes using multipartite continuous-variable entangled states. This network setup is a novel scheme that has not been investigated previously. The trade-off in this proposed scheme of N-connected nodes is the attained teleportation lengths that are reduced because of the corresponding Bell measurements (which weaken the CV-entanglement).

The remainder of the paper is organized as follows. In Section 2, the proposed system is described, and the governing Hamiltonian is presented. In Section 3, the equation of motion is first derived, followed by a stationary covariance matrix that describes the system stability. Multipartite entanglement is modeled and evaluated in Section 4. In Section 5, the teleportation network and numerical estimations of the transfer fidelity are discussed. Finally, conclusions are drawn in Section 6.
\begin{figure}[h!]
\centering\includegraphics[width=7cm]{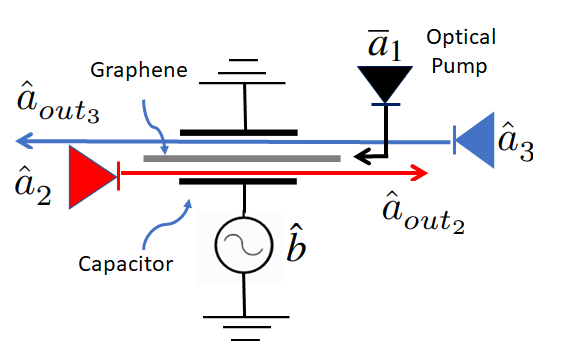} 
\caption{ The superconducting graphene-loaded capacitor driven by a nonclassical microwave field and pumped by a classical optical pump as an SPP graphene mode to compensate for losses. Two interacting quantum SSP fields counterpropagating along the graphene waveguide that are electrically coupled to the same microwave fields.} 
\label{fig12}
\end{figure}
\section{System}
The basic building unit of the proposed system is our previously reported graphene-loaded capacitor \cite{asjad21,qasIeee, qasUS.2}, which consists of a graphene plasmonic layer integrated with a parallel-plate electrical capacitor, as illustrated in Fig 1. The graphene layer occupies the $yz$-plan and supports propagating optical surface plasmon polariton (SPP) modes. The electric field of the SPP propagating mode is given by $E_{sp}=A(y,z) D(x)e^{-i(\omega t -\beta z)}$, where $A(y,z)$ is the complex amplitude, $i=\sqrt{-1}$, and $D(x)$ is the exponentially decaying spatial transverse profile. A microwave signal, $V_m=\nu e^{-i\omega_m t}$+c.c., is driving the electrical capacitor. Consequently, the graphene chemical potential is given by $\mu_c=\hbar v_F\sqrt{\pi n_0+2CV_m/(A_r e)}$. Here, $n_0$ and $C$ are the intrinsic electron density and the capacitance per unit area, respectively, $A_r$ is the electrodes' cross-sectional area, and $v_F$ is the fermi velocity of the Dirac fermions.

For small applied electric microwave voltages, the chemical potential and the graphene conductivity are expanded to the first order, yielding $\mu_c=\mu^{(1)}_c+\nu \mu^{(2)}_c e^{-i\omega_m t}+c.c.$ and $\sigma_s= \sigma_s^{(1)}+\nu \sigma_s^{(2)} e^{-i\omega_m t}+c.c.$ (see Appendix A for $\sigma_s^{(1)}$ and $\sigma_s^{(2)}$ expressions). The dispersion relation of the SPP mode is given by $\beta=k\sqrt{1-4/(Z_0 \sigma_s)^2}$, where $k=\omega/c$ stands for the propagation constant, and $Z_0=377\Omega$ denotes the free space impedance. Similarly, the propagation coefficient $\beta$ and the effective permittivity $\epsilon_{eff}$ of the SPP mode are described by first order expansion, reading $\beta= \beta ^{(1)}+\nu \beta^{(2)} e^{-i\omega_m t}+c.c.$, and $\epsilon_{eff}=\epsilon^{(1)}_{eff}+\nu \epsilon^{(2)}_{eff} e^{-j\omega_m t}+c.c.$, respectively. Here, $\mu^{(1)}_c=\hbar v_F \sqrt{\pi n_0}$, $\mu^{(2)}_c=\hbar v_F C/e\sqrt{\pi n_0}$, $\epsilon^{(1)}_{eff}=(\beta^{(1)}/k_0)^2$, $\epsilon^{(2)}_{eff}=2\beta^{(1)}\beta^{(2)}_i/k_i^2$,   $\beta^{(2)}=\beta^{(1)}\sigma^{(2)}_s/[\sigma^{(1)}_s(1-(Z_0 \sigma^{(1)}_s/2)^2)]$, and $c.c.$ remarks complex conjugate.

Three SPP modes with distract frequencies (i.e., $\omega_1$, $\omega_2$ and $\omega_3$) are considered co-propagating along with the graphene layer. The corresponding classical Hamiltonian is given by $\mathcal{H} = \frac{1}{2} \nu^2 C +\frac{1}{2} \iiint_{x,y,z}\big( \varepsilon_{0}\varepsilon_{eff}\lvert \vec{E}_t\rvert\ ^{2} +\mu_0\lvert \vec{H}_t\rvert\ ^{2} \big) \partial x \partial y  \partial z$, where $\vec{E}_{t}=\vec{E}_{SP_1}+\vec{E}_{SP_2}+\vec{E}_{SP_3}$ is the total SPP electric field, $\vec{E}_{SP_{1,2,3}}$ are the electric fields of the SPP modes at the corresponding frequencies $\omega_{1,2,3}$, respectively, and $\vec{H}_t$ is the total SPP magnetic field. It is important to note that owing to the perturbed effective permittivity presented above, the three SPP modes and the  driving microwave field are coupled under the condition $\omega_m=\omega_3-\omega_1=\omega_2-\omega_2$. The SPP mode at $\omega_1$ is implemented as a classical pump signal to compensate for the propagation losses. However, the other two SPP modes at $\omega_2$, $\omega_3$, and the microwave field at $\omega_m$ are quantum fields. 

In this study, we consider an array (of chains) of  $\mathcal{N}$-independent plasmonic graphene-loaded capacitors, as shown in Fig. 2(a). The quantum Hamiltonian can be obtained from its classical counterpart by quantizing the interacting fields through the relations $A_ \iota= \big(\hbar \omega_ \iota\big)^{\frac{1}{2} }\bigg(\varepsilon_{0}\varepsilon_{eff_{ \iota}}^{(1)} V_L \bigg)^{-\frac{1}{2}}\hat{a}_ \iota$, and $ \nu= \bigg( \frac{2\hbar \omega_m}{ C \mathcal{A}_r}\bigg)^{\frac{1}{2}} \hat{b}$. Here, $\iota\in \{1,2\}$, $V_L$ is the volume of the SPP mode. Consequently, the governing quantum Hamiltonian $\hat{H}=\hat{H}_0+\hat{H}_I$ reads \cite{asjad21}:
\begin{equation}\label{eq1}
\hat{H}_0=\sum^N_j \omega^j_m \hat{b}^{j \dagger} \hat{b} +\omega^j_1 \bar{a}^{j^ *}_1 \bar{a}^j_1+\omega^j_2 \hat{a}^{j^ \dagger}_2 \hat{a}^j_2 +\omega^j_3 \hat{a}^{j^ \dagger}_3 \hat{a}^j_3, 
\end{equation} and 
\begin{equation}\label{eq2}
\hat{H}_I =\sum^N_j g^j_2 (\hat{a}^{j\dagger}_2 \, \bar{a}^j_1 \, \hat{b}^j+  \bar{a}^{j *}_1  \, \hat{a}^j_2 \, \hat{b}^{j\dagger} )+g^j_3 (\bar{a}^{j*}_1\, \hat{a}^j_3\, \hat{b}^j+ \hat{a}^{j\dagger}_3 \, \bar{a}^j_1 \,\hat{b}^{j\dagger}).
\end{equation}
Here, $j \in \{1,2,3,...,N\}$, $g^j_{2,3}$ denotes the perturbation coupling coefficients \cite{asjad21}, and the condition $\omega^j_{m} =\omega^j_2-\omega^j_1=\omega^j_1-\omega^j_3$ is satisfied for every $j$-th element. The principle of the building unit modality is based on coupling the quantum optical fields (of annihilation operators $\hat{a}^j_2$ and $\hat{a}^j_3$ of $\omega_1$ and $\omega_2$ frequencies, respectively) which co-exist as counter copropagating surface plasmon polariton (SPP) modes along with the graphene layer. The microwave voltage (of annihilation $b^j$ at frequency $\omega_m$) drives the capacitor, enabling the interaction process by electrically perturbing the graphene conductivity \cite{qsasexpress,qasUS1}. Losses are compensated by launching a copropagating intense SSP optical pump (with an amplitude $\bar{a}^j_1$ and a frequency $\omega_1$) \cite{QasSpring}.
\begin{figure}[h!]
\centering\includegraphics[width=1\columnwidth]{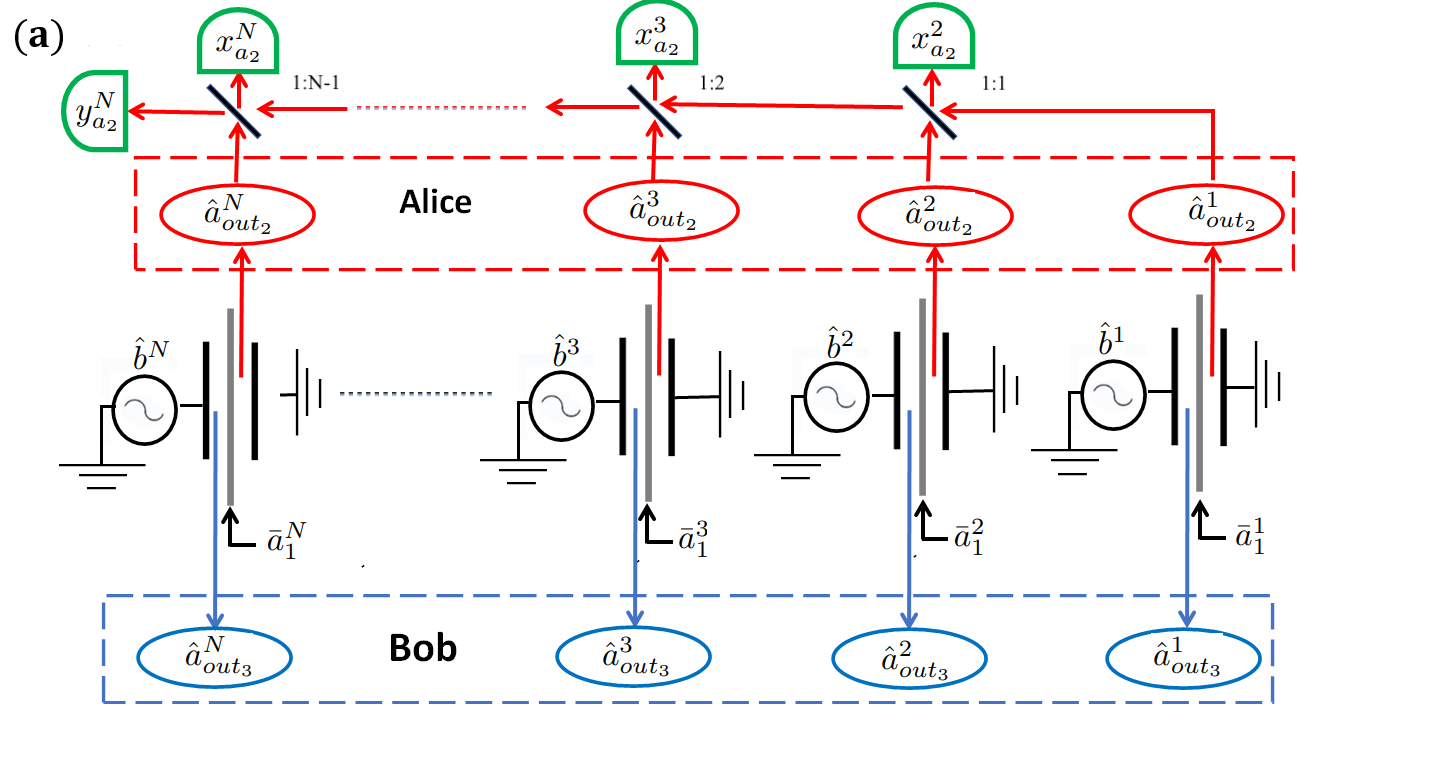} 
\centering\includegraphics[width=0.5\columnwidth]{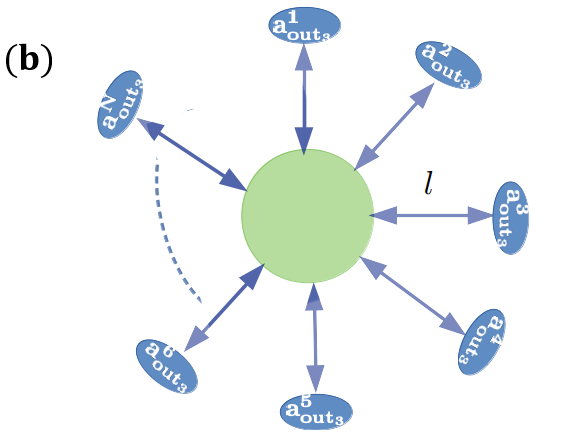}
\caption{ (a) The structure of the proposed teleportation network composed of an array of graphene-loaded superconducting capacitors. The paired outputs of the elements (Alice and Bob modes) are processed using a system of linear optical components (beam splitters and detectors) to achieve bipartite entanglement among the Bob modes on demand. (b) An effective star-like network whereby bipartite entanglement can be produced on demand between any pair ($\hat{a}^i_{out_3}$, $\hat{a}^j_{out_3}$) of Bob modes.} 
\label{fig1}
\end{figure}

\section{Equation of Motion} 
SPP pump modes with amplitudes $\bar{a}^j_1$ at a frequency $\omega_1$ are considered intense and can be treated classically. The Hamiltonian given by Eq. (1) and Eq. (2) can thus be used to obtain the equations of motion for the microwave and SPP modes of the $j$-th element in the context of open system dynamics under the Heisenberg-Langevin formalism given below:
\begin{equation}
 \label{141}
\dot{\hat{b}}^j=-\gamma^j_m \hat{b}^j-i\mathcal{G}^j_2 \hat{a}^j_2 -i\mathcal{G}^j_3 \hat{a}^{j \dagger}_3 +\sqrt{2 \gamma^j_m}\, \hat{b}^j_{in},
\end{equation}

\begin{equation}
\label{1412}
\dot{\hat{a}}^j_2=-\gamma^j_2 \hat{a}^j_2 - i\mathcal{G}^j_2 \hat{b}^j +\sqrt{2 \gamma^j_2}\, \hat{a}^j_{in_2},
\end{equation}

\begin{equation}
\label{1413}
\dot{\hat{a}}^j_3=-\gamma^j_3 \hat{a}^j_3 - i\mathcal{G}^j_3 \hat{b}^{j \dagger} +\sqrt{2 \gamma^j_3} \, \hat{a}^j_{in_{3}}, 
\end{equation}
where $\mathcal{G}^j_2= \bar{a}^j_1 g^j_2$, $\mathcal{G}^j_3= \bar{a}^j_1 g^j_3$, and $\gamma_m$, $\gamma^j_2$ and $\gamma^j_3$ are the decay rates of the microwave and SPP modes, respectively. Here, $\hat{a}^j_{in_2}$ and $\hat{a}^j_{in_3}$ are the input noise operators, characterized by $\langle \hat{a}_{in_2}(t) \hat{a}^\dagger_{in_2}(t')\rangle=\delta(t-t')$ and $  \langle \hat{a}_{in_3}(t) \hat{a}^\dagger_{in_3}(t')\rangle=\delta(t-t') $, respectively. The equations in (3) to (5) are obtained for a rotating frame at $\omega_m$, $\omega_2$ and $\omega_3$. Furthermore, we consider a nonclassical driving microwave voltages with single-mode Gaussian states. The associated noise operator $\hat{b}^j_{in}$ is characterized by $\langle \hat{b}^j_{in}(t) \hat{b}^{j}_{in}(t')\rangle=M^j \exp^{i\phi}\delta(t-t')$ and $\langle \hat{b}^j_{in}(t) \hat{b}^{j^\dagger}_{in}(t')\rangle=(N^j+1) \delta(t-t')$. Here, $M^j$ and $N^j$ are given by \cite{Tahira}:
\begin{eqnarray}
M^j =\dfrac{1}{4\mathcal{P}^{j^2}(1-2\mathcal{D}^j)}-\dfrac{1-2 \mathcal{D}^j}{4},\\
N^j =\dfrac{1}{4\mathcal{P}^{j^2}(1-2\mathcal{D}^j)}+\dfrac{1-2 \mathcal{D}^j}{4}.
\end{eqnarray}
where $\mathcal{D}^j$ and $\mathcal{P}^j$ are the {\it purity}  and {\it nonclassicality depth}. The degree of mixedness of the prepared quantum state is characterized by the purity $\mathcal{P}^j$, which ranges from $0$ to $1$ (with $0$ being for mixed states and 1 being for pure states), while the nonclassicality depth $\mathcal{D}^j$ ranges from $0$ to $0.5$.
For the sake of simplicity and without losing any generality, we assume that all microwave modes have the same {\it purity}  $\mathcal{P}^j=\mathcal{P}$ and {\it nonclassicality depth}  $\mathcal{D}^j=\mathcal{D}$. Consequently, and in accordance with input-output theory \cite{gardiner85}, the output field operators $ \big( \hat{a}^j_{out_2}$ and $\hat{a}^j_{out_3}\big)$ are related to the corresponding input operators $\big(  \hat{a}^j_2$ and $\hat{a}^j_3 \big) $ by $\hat{a}^j_{out_2}(t)=\sqrt{\gamma^j_{2}}  \hat{a}^j_2(t) - \hat{a}^j_{in_2}(t) $ and $\hat{a}^j_{out_3}(t)=\sqrt{\gamma^j_3} \hat{a}^j_3(t) - \hat{a}^j_{in_3}(t) $, respectively.
It then follows that the density matrix $\hat{\rho}^{out}_{\bf a_2 a_3}$ of the output field modes, ${\bf \hat{a}}_{\mathcal{F}, out_2} (t)=\Big( \hat{a}^1_{out_2} (t), \hat{a}^2_{out_{2}} (t),..,\hat{a}^N_{out_{2}} (t)\Big)$, named Alice modes, and ${\bf \hat{a}}_{out_3} (t)=\Big(\hat{a}^1_{ out_{3}} (t), \hat{a}^2_{ out_{3}} (t),..,\hat{a}^N_{out_{3}} (t) \Big)$, named Bob modes, can be expressed as $\hat{\rho}^{out}_{\bf a_2 a_3}= \hat{\rho}^{1 \, out}_{a_2 a_3} \otimes  \hat{\rho}^{2\, out}_{a_2 a_3} ,.....,\otimes  \hat{\rho}^{N \,out}_{a_2 a_3}$, where
\begin{equation}
\hat{\rho}^{j\,out}_{a_2 a_3}= \int \mathcal{\chi}_{j}(\bf{ \varepsilon}^j_{a_2},\varepsilon^j_{a_3}) \mathcal{D}(\varepsilon^j_{a_2}) \otimes \mathcal{D}^\dagger(\varepsilon^j_{a_3}) d^2 \varepsilon^j_{a_2} d^2 \varepsilon^j_{a_3}.
\end{equation}
Here, $\mathcal{D}(\varepsilon^j_{a_2})=\exp(\varepsilon^j_{a_2} \hat{a}^{j \dagger}_{out_{2}} - \varepsilon^{j*}_{a_2} \hat{a}^j_{out_{2}} )$ and $\mathcal{D}(\varepsilon^j_{a_3})=\exp(\varepsilon^j_{a_3} \hat{a}^{j \dagger}_{out_{3}}- \varepsilon^{j*}_{a_3} \hat{a}^j_{out_{3}} )$ are the displacement operators for the $\textit{j-th}$ Alice and Bob modes, respectively. The complex phase-space variables $\varepsilon^j_{a_2} = (x^j_{a_2} + i y^j_{a_2})/\sqrt{2}$ and $\varepsilon^j_{a_3} =(x^j_{a_3} + i y^j_{a_3})/\sqrt{2}$ correspond to the bosonic operators $\hat{a}^j_{out_{2}}=(\hat{x}^j_{a_2} + i \hat{y}^j_{a_2})/\sqrt{2}$ and $\hat{a}^j_{out_{3}}=(\hat{x}^j_{a_3} + i \hat{y}^j_{a_3})/\sqrt{2}$, respectively. The real phase-space variables $x^j_{a_2}$ ($x^j_{a_3}$) and $y^j_{a_2}$ ($y^j_{a_3}$) are the counterparts of the Hermitian quadrature operators $\hat{x}^j_{a_2}$ ($\hat{x}^j_{a_3}$) and $\hat{y}^j_{a_2}$ ($\hat{y}^j_{a_3}$), respectively. Furthermore, $\chi(\bf{ \varepsilon}_{a_2},  \bf{\varepsilon}_{a_3})$ is the characteristic function, which is the Fourier transform of the Wigner function, i.e., $\chi(\bf{ \varepsilon}_{a_2},\varepsilon_{a_3}) =\mathcal{F}\left[ \mathcal{W}({\bf \varepsilon}_{a_2}, \varepsilon_{a_3})\right]$.
The joint Wigner function for $2 N$ output modes can be formalized for zero-mean Gaussian quantum states as
\begin{equation}
\mathcal{W}({\bf r_{a_2}},{\bf r_{a_3}})= \mathcal{N}_r \exp \left\{-\dfrac{1}{2} ({\bf r_{a_2}}, {\bf r_{a_3}})\mathcal{V}^{-1} ({\bf r_{a_2}},{\bf r_{a_3}})^T\right\},
\end{equation}
where $\mathcal{N}_r$ is the normalization factor, the subscript$^T$ denotes the transpose, ${\bf r_{a_2}}=(x^1_{a_2}, y^1_{a_2}, x^2_{a_2}, y^2_{a_2},$ $..., x^\mathcal{N}_{a_2}, y^\mathcal{N}_{a_2})$ and ${\bf r_{a_3}}=(x^1_{a_3}, y^1_{a_3}, x^2_{a_3}, y^2_{a_3}, ..., x^\mathcal{N}_{a_3}, y^\mathcal{N}_{a_3})$ are the vectors of the real variables associated with the Alice and Bob modes, and $\mathcal{V}$ is the $4 N \times 4 N$ covariance matrix (CM), which can be written in the following block form:
\begin{equation}
\mathcal{V}=\begin{pmatrix}
\mathcal{V}_{a_2} \mathcal{I}_{2N}&  \mathcal{V}_{a_2a_3} \mathcal{I}_{2N}\\
\mathcal{V}^T_{a_2 a_3} \mathcal{I}_{2N}&  \mathcal{V}_{a_3} \mathcal{I}_{2N}\\
\end{pmatrix}.  
\end{equation}
Here, $\mathcal{V}_{a_2}=\mathrm{diag}({\mathcal{V}^1_{a_2},\mathcal{V}^2_{a_2},\mathcal{V}^3_{a_2},..., \mathcal{V}^N_{a_2} }) $ and $\mathcal{V}_{a_3} =\mathrm{diag}({\mathcal{V}^1_{a_3},\mathcal{V}^2_{a_3},..., \mathcal{V}^N_{a_3} })$ are the $2N \times 2N$ covariance matrices for the Alice and Bob modes, respectively. Additionally, $\mathcal{V}_{a_2a_3}$ is a $2N \times 2N$ matrix that describes the correlation between the Alice $(\hat{r}_a)$ and Bob $(\hat{r}_{a_3})$ modes. The stationary covariance matrix $\mathcal{V}^j$ for the {\it j-th} pair of Alice and Bob modes is given by
\begin{equation}
\mathcal{V}^j=\int^\infty_{-\infty}  \mathcal{Q}^j  \mathcal{T}^j(\omega) \mathcal{N}^j_{in} \mathcal{T}^{jT}(-\omega) \mathcal{Q}^{j T}  d\omega,
\end{equation}
where $\mathcal{N}^j_{in}=\text{diag}(\mathcal{N}^j_2, \mathcal{N}^j_3, \mathcal{N}^j_b)$ is the diffusion matrix, $\mathcal{N}^j_2=\mathcal{N}^j_3=\begin{pmatrix}
0 &   1\\
0 &0
\end{pmatrix} $, $\mathcal{Q}^j =\text{diag} ( \mathcal{Q}^j_2, \mathcal{Q}^j_3, \mathcal{Q}^j_b)$, $\mathcal{Q}^j_3=\dfrac{1}{2}  \begin{pmatrix}
1 &   1\\
-i  & -i
\end{pmatrix} $, and $\mathcal{N}^j_b=\begin{pmatrix}
M^j&   N^j+1\\
N^j_m &M^*
\end{pmatrix} $.  Here, $\mathcal{T}^j(\omega)= F^j(\omega) [\mathcal{A}^j-i\omega]^{-1}\nu^j-{\bf I}$, $\nu=\text{diag}(\sqrt{2\gamma^j_2},\sqrt{2\gamma^j_2},\sqrt{2\gamma^j_3},\sqrt{2\gamma^j_3} ,\sqrt{2\gamma^j_m},\sqrt{2\gamma^j_m} )$, $F^j(\omega)= \text{diag}(\sqrt{2\gamma^j_2},\sqrt{2\gamma^j_2} ,\sqrt{2\gamma^j_3}, \sqrt{2\gamma^j_3} ,1,1)$, and the drift matrix for the \textit{j-th} element in the $\mathcal{A}^j$ array is given by
\begin{equation}
\mathcal{A}^j=\begin{pmatrix}
-\gamma^j_2 &  0& 0& 0 & -i \mathcal{G}^j_2&0 \\
 0 & -\gamma^j_2 &0 & 0& 0 & i \mathcal{G}^j_2 \\
 0 &   0              & -\gamma^j_3 & 0 & 0& -i \mathcal{G}^j_3 \\
 0 & 0&0 & -\gamma^j_3  & i \mathcal{G}^j_3 &0 \\
 -i \mathcal{G}^j_2 &   0    & 0 & -i \mathcal{G}^j_3 & -\gamma^j_m& 0 \\
 0 & i \mathcal{G}^j_2&i \mathcal{G}^j_3 & 0  &0 &-\gamma^j_m \\
\end{pmatrix}. 
\end{equation}
According to the Routh-Hurwitz criterion \cite{deJesus87}, the stability of the steady-state solution can be guaranteed if the real part of the eigenvalues of $\mathcal{A}^j$ are negative. In this work,  proper parameters are considered to attain stable solutions (see the stability analysis in the Appendix B).

\section{Multipartite Entanglement}
In this section, we show how N-distant independent plasmonic graphene waveguides can be used to generate an N-partite CV entangled state. This system can be realized by sending the output of the Alice modes ($ \hat{a}^j_{out_2}$) from each plasmonic graphene waveguide to an intermediate 
conman 
node (named Charlie), where multipartite Bell measurement is performed. An N-partite entangled state of the Bob modes ($ \hat{a}^j_{out_3}$) is thus prepared. As shown in Fig. 2(a), Charlie combines the Alice modes on an array of $N-1$ 
beam splitters (BS) with respective ratios of 1:1, 1:2, ..., 1 : N-1 and then performs multipartite homodyne detection on the BS output fields. The classical result $\bar{r}_a$ is generated. Accordingly, the transformed position (phase) quadratures of the Alice modes are given by \\
\begin{equation}
\hat{x}^1_{a_2}\rightarrow \sum^N_j\hat{x}^j_{a_2}/\sqrt{N}\,\,\,\,\, (\hat{y}^1_{a_2}\rightarrow \sum^N_j\hat{y}^j_{a_2}/\sqrt{N}),
\end{equation}

\begin{equation}
\begin{split}
&\hat{x}^{N-1}_{a_{2}}\rightarrow\left(\sum^{N-1}_j \hat{x}^j_{a_2} - (N-1) \hat{x}^N_{a_2} \right)/\sqrt{N (N-1)}
\\&\left(\hat{y}^{N-1}_{a_{2}} \rightarrow\left\{\sum^{N-1}_j \hat{y}^j_{a_2} - (N-1) \hat{y}^N_{a_2} \right\}/\sqrt{N (N-1)} \right).
\end{split}
\end{equation}
Note that the first $N-1$ outputs $\hat{a}^j_{\mathcal{F}, out_2}$ (with $j=1,2,3, .., N-1$) in the multipartite Bell measurement of the Alice modes are homodyne when detected in the position quadrature $\hat{x}^j_{a_2}$, whereas the last output $\hat{a}^N_{\mathcal{F}, out_2}$ is detected in the phase quadrature $\hat{y}^N_{a_2}$. As a result, all the Bob  modes can be efficiently driven into an N-partite entangled state. The corresponding Wigner function of the conditioned Bob' modes for the detection result $\bf \bar{r}_a$ reads
\begin{equation}\label{101}
\mathcal{W}({\bf r_b/{\bf \bar{r}_a}})= \mathcal{N}' \exp \left\{-\dfrac{1}{2}  {\bf r_b} \mathcal{V'}^{-1} {\bf r_b}^T \right \},
\end{equation}
where $\mathcal{V'}$ is a $2 N \times 2 N$ covariance matrix that describes the N-partite entangled Gaussian state. Note that first-order terms are not shown in Eq. (\ref{101}) because these terms have a negligible impact and can zeroed by considering appropriate feedback. The covariance matrix  $\mathcal{V'}$ for N-identical plasmonic graphene waveguides ($\mathcal{V}_{a^j_2} =\mathcal{V}_{a}$, $\mathcal{V}_{a^j_3} =\mathcal{V}_{a_3}$ and $\mathcal{V}_{a^j_2 a^j_3} =\mathcal{V}_{a_2 a_3}$) can be written in the following block form:
\begin{equation}
\mathcal{V'}=\begin{pmatrix}
\mathcal{V}'_{a_3}     & \mathcal{V}'_{a_3a_3} &  \cdots        & \mathcal{V}'_{a_3a_3}             \\
\mathcal{V}'_{a_3a_3}   &\mathcal{V}'_{a_3}    &                   &   \mathcal{V}'_{a_3a_3}             \\
\vdots                       &                                & \ddots       &            \vdots                                      \\
\mathcal{V}'_{a_3a_3}    &        \cdots                        &      \mathcal{V}'_{a_3a_3}             &     \mathcal{V}'_{a_3}  &    
\end{pmatrix},  
\end{equation} 
where $\mathcal{V}'_{a_3} = \mathcal{V}_{a_3} - (N-1)\mathcal{V}_{a_2 a_3}  \mathcal{Z}_1   \mathcal{V}^{-1}_{a_2}  \mathcal{Z}_1   \mathcal{V}_{a_2 a_3}/N - \mathcal{V}_{a_2 a_3}  \mathcal{Z}_2 \mathcal{V}^{-1}_{a_2}  \mathcal{Z}_2   \mathcal{V}_{a_2 a_3}/N$ and $ \mathcal{V}'_{a_3 a_3} =\mathcal{V}_{a_2 a_3}  \mathcal{Z}_1   \mathcal{V}^{-1}_{a_2}  \mathcal{Z}_1   \mathcal{V}_{a_2a_3}/N- \mathcal{V}_{a_3a_3}  \mathcal{Z}_2 \mathcal{V}^{-1}_{a_2}  \mathcal{Z}_2 \mathcal{V}_{a_2 a_3}/N$ are $2 \times 2$ submatrices with $\mathcal{Z}_1 =\mathrm{diag} (1, 0)$ and  $\mathcal{Z}_2 =\text{diag} (0, 1)$.

The stationary entanglement between any pair of Bob modes can be measured by the logarithmic negativity \cite{Plenio, Adesso, Vidal2002}:
\begin{equation}\label{En}
E^{(j)}_{\mathcal{N}}=\max[0, -\ln 2 \eta^-_j],
\end{equation}
where  $\eta^-_j$ 
is the smallest symplectic eigenvalue of the partially transposed covariance matrix $\mathcal{V'}^j$ of the {\it j-th} pair of Bob modes. A nonzero value of $E^{(j)}_\mathcal{N}$ can be used to quantify the degree of entanglement between the {\it j-th} pair of Bob modes.

The N-partite stationary entanglement at the output can be exploited to realize a quantum network. The channel- and transmission-associated losses can be described using the concept of an effective beam splitter with a transmissivity $\eta=\eta_0e^{-\alpha l/10}$ \cite{barbosa11}, where $\alpha$ is the classical channel attenuation in $dB/km$, $\eta_0$ describes all possible inefficiencies, and $l$ is the distance traveled by each field (the classical channel length) \cite{Asjad16, Asjad15}. The optical outputs are transmitted by free space channels. The associated attenuation is attributed to the free space scattering and absorption \cite{Glickman}. On a clear day, the visibility is between $50-150$ km. The corresponding attenuation parameter varies between $0.04$ and $0.005$ dB/km for optical wavelengths between 850 nm and 1550 nm, respectively \cite{Fischer}. We note here that typical techniques used in wireless systems can be adapted to compensate for fading and other random variables parameters that are attributed to the stochastic nature of the free-space channels. However, these scenarios are beyond the scope of this work, and our calculations are carried out using average values of the attenuation parameters. The developed model in this study lay the ground for future investigations that can be dedicated to characterizing stochastic free space channels for teleportation networks.
The corresponding output covariance matrix is given by  $\mathcal{V'}_{los}= \eta \mathcal{V'}+\dfrac{1}{2}(1-\eta) {\bf I}$, where $ {\bf I}$ is the $2N\times 2N$ identity matrix. Here, all the Bob modes are assumed to be equidistant from the central hub (i.e., $l$).  
\begin{figure}[h!]
\centering\includegraphics[width=1\columnwidth]{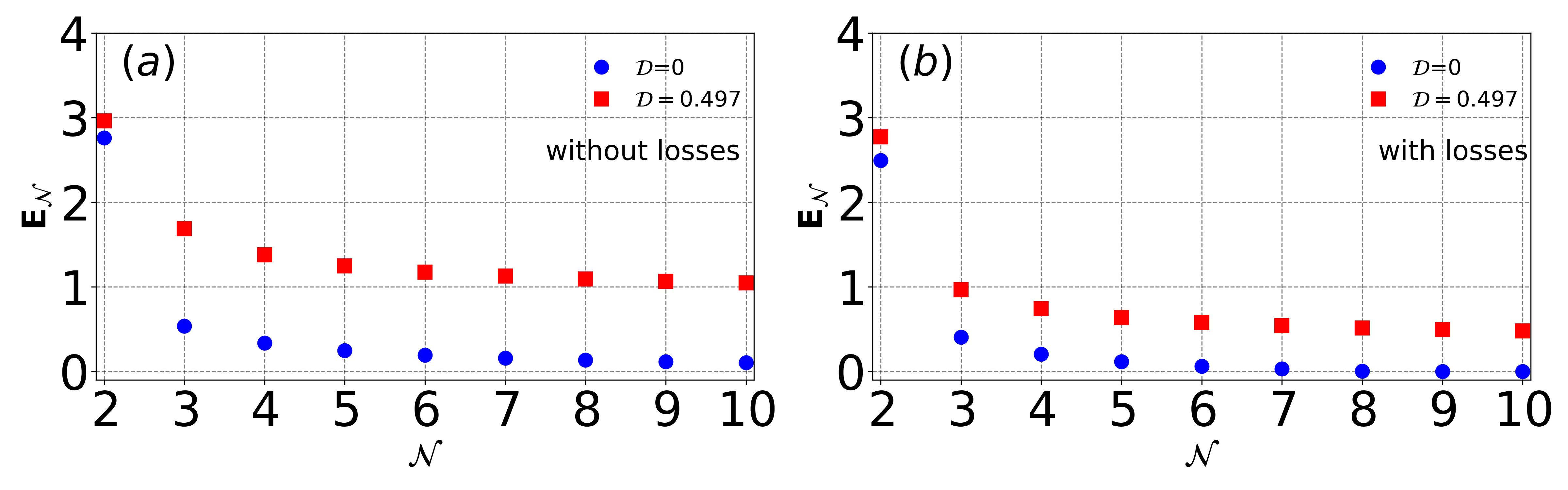} 
\caption{ Logarithmic negativity $E_\mathcal{N}$ between any pair of Bob modes as a function of the number of modes $N$ in the limit of zero bandwidth. Two cases of nonclassical depth $\mathcal{D}=0$ (blue dots) and $\mathcal{D}=0.497$ (red squares) are considered. In (a), losses are neglected, whereas in (b), a realistic channel of free space (with $\alpha=0.005$) and a detection efficiency  $\eta_{0}=99\%$ are considered. The other parameters are $\gamma_m/\omega_m=0.001$, $\gamma_2/ \omega_m=\gamma_3/ \omega_{m}=0.02$, $\mathcal{G}_{2} / \omega_m=0.2$, $\mathcal{G}_{3} / \omega_m=0.14$, and $l=0.1 km$. } 
\label{fig2}
\end{figure}

Fig. (\ref{fig2}) shows the calculated logarithmic negativity for the proposed scheme versus the number of Bob modes (the number of independent plasmonic waveguides). The calculations are performed for different degrees of nonclassicity of the driving microwave field. Additionally, the simulations are performed in the absence and presence of losses, as shown in Fig. \ref{fig2}(a) and Fig. \ref{fig2}(b), respectively. The nonclassicality associated with the microwave field enables an increase in the entanglement between any pair of Bob modes (red squares). For instance, for the parameters considered in Fig. (\ref{fig2}) of $\omega_1=193$ THz, $\gamma_m/ \omega_m=0.001$, $\gamma_2/ \omega_m=\gamma_3/ \omega_m=0.02.$, $\mathcal{G}_2/ \omega_m=0.2$, $\mathcal{G}_3/ \omega_m=0.14$, $\eta_{0}=0.99$ and $l=0.1 km$ and $\alpha=0.005$, our numerical investigations show that the logarithmic negativity is maximally boosted for $\mathcal{D}=0.497$ (see Appendix C). However, when the number of Bob modes increases, the logarithmic negativity becomes more sensitive (degrades). Our simulations show  that incorporating the nonclassicality of the driving microwave field results in a logarithmic negativity greater than zero for up to 10 Bob modes with separation distances of approximately $0.1$ km. Hence, it is demonstrated that Charlie can generate any type of entangled state among different numbers of Bob modes by choosing a proper array of BSs.
\begin{figure}[h!]
\centering\includegraphics[width=1\columnwidth]{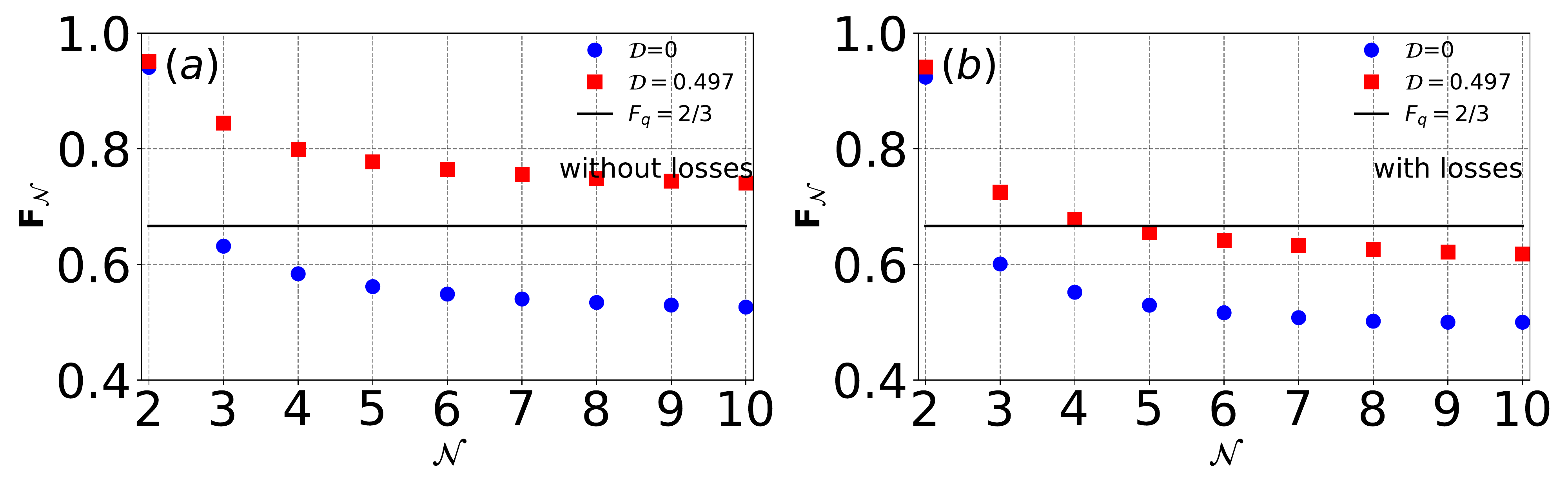} 
\caption{Optimal teleportation fidelity $F_{\mathcal{N}}$ as a function of the number of Bob modes. In (a), losses are neglected, whereas in (b), a realistic channel of free space (with $\alpha=0.005$) and a detection efficiency  $\eta_{0}=99\%$ are considered. All parameters are the same as those shown in Fig. (\ref{fig2}). The horizontal black line corresponds to the secure quantum teleportation threshold $F_{q}=2/3$. }
\label{fig3}
\end{figure}

\section{ Teleportation Network}
The output N-partite Gaussian entangled states can be characterized and function 
as a quantum channel for multipartite quantum teleportation. To this end, we analyze the performance of this multipartite quantum channel realized by the Bell measurements in terms of the teleportation fidelity of a pure coherent state among the Bob modes ($\hat{a}^1_3, \hat{a}^2_3,\hat{a}^3_3,...., \hat{a}^N_{3}$). Therefore, for CV teleportation protocols, Bob combines an unknown input coherent state $|\alpha_{in}\rangle$ (that is to be teleported) with the part of the entangled state in his hand, $\hat{a}^1_3$, on a beam splitter and measures two quadratures $1/\sqrt{2} (\hat{x}_{in}-\hat{x}^1_{a_3})$ and $1/i \sqrt{2}(\hat{p}_{in}-\hat{p}^1_{a_3})$, where $\alpha_{in} = (\hat{x}_{in} + i \hat{p}_{in})/\sqrt{2}$. The measurement outcomes are sent to $N-1$ receivers simultaneously. Each of the $N-1$ Bob modes displaces its state according to the measurement outcomes. The corresponding optimal teleportation fidelity is given by \cite{mari08}
\begin{equation}
F_{a^1_3:a^j_3}=\dfrac{1}{1+2 \eta^-_j},
\end{equation}
where $\eta^-_j$ is equivalent to the smallest symplectic eigenvalue of the partially transposed $\mathcal{V}'$ under the bipartition $a^1_3: a^2_3a^3_3 \cdots a^{N-1}_3$. The optimal fidelity is directly related to the logarithmic negativity $E^{(n)}_{\mathcal{N}}$.

In Fig. (\ref{fig3}), the optimal teleportation fidelity $F_\mathcal{N}$ of the unknown coherent state is calculated as a function of the number of Bob modes. Here, zero and $\mathcal{D}=0.497$ nonclassicality depths are considered. Additionally, the cases of lossless and realistic free space channels are evaluated in Fig. \ref{fig3}(a) and Fig. \ref{fig3}(b), respectively. The calculations show that the nonclassicality of the driving microwave field enables the threshold for a secure quantum teleportation limit $F_{q} = 2/3$ (the black horizontal line) to be exceeded, even in the presence of losses. However, our simulations show that up to only four Bob modes have fidelity beyond the threshold.
\begin{figure}[h!]
\centering\includegraphics[width=1\columnwidth]{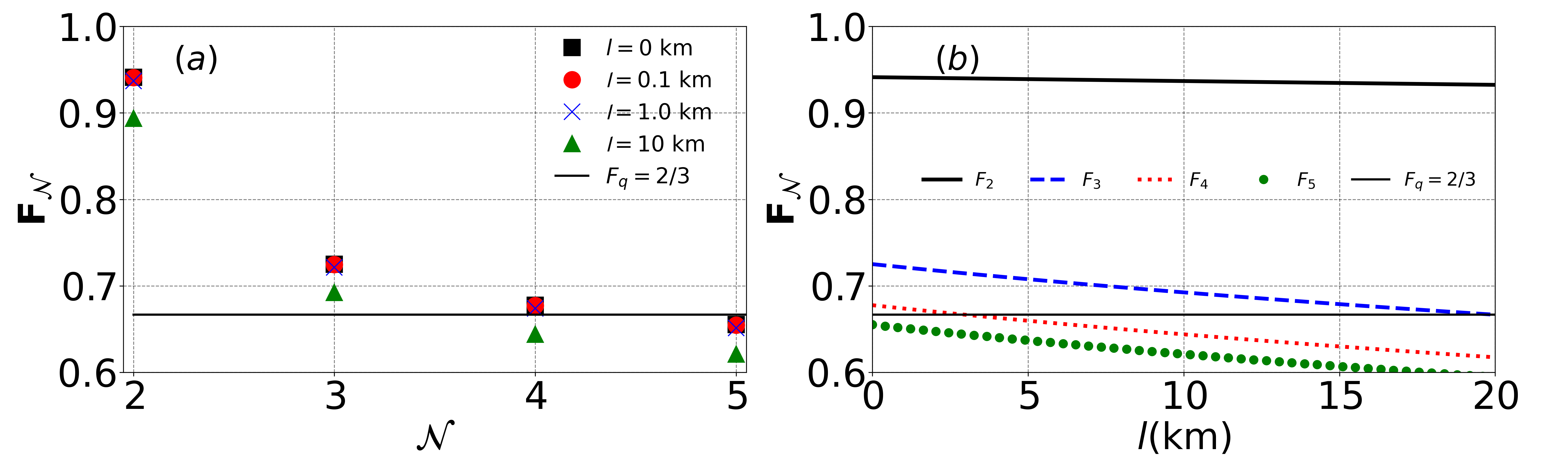} 
\caption{ (a) Optimal teleportation fidelity $F_{\mathcal{N}}$ as a function of the number of Bob modes. Different values of the distance $l=0.0$ km (black), 0.1 km (red), 1.0 km (blue), 10 km (green) are considered. (b) Optimal teleportation fidelity $F_{\mathcal{N}}$ as a function of transmission length $l$ for different number of Bobs mode. The optimal nonclassicality depth $\mathcal{D}=0.497$ and all other parameters are the same as those shown in Fig. (\ref{fig2}). }
\label{fig4}
\end{figure}

The robustness of the teleportation fidelity versus the number of Bob modes $\mathcal{N}$ is displayed in Fig. \ref{fig4} (a). Different values of the separation distance $l$ are considered. Here, $\mathcal{D}=0.497$, and realistic free space channel losses are taken into account ($\alpha=0.005$). All other parameters are the same as those shown in Fig. (\ref{fig2}). It can be inferred that significant fidelity greater than the threshold $2/3$ can be achieved for the proper combination of the number of Bob modes and channel transmission lengths. In Fig. \ref{fig4}(b), the optimal teleportation fidelity $F_{\mathcal{N}}$ is calculated as function of transmission length $l$ for different number of Bob modes. 
Interestingly, the teleportation fidelity can bear the required quantum threshold for $\mathcal{N} = 3$ and $l = 20 km$ (blue dashed curve) and alternatively for $\mathcal{N} = 4$ and $l =2.3 Km$ (red dotted curve). However, for a larger number of Bob modes such as $\mathcal{N}=5$ (red dotted curve), the secure teleportation condition is not satisfied. As a consequence, there is a trade-off between the separation distance and the number of Bob modes. This very promising result demonstrates the feasibility of realizing quantum teleportation network based on the proposed system.
\section{Conclusion}
We have presented a novel scheme for generating multipartite continuous-variable entangled states among remotely connected independent nodes. The proposed system employs an array of separated graphene plasmonic waveguides that are activated by biasing nonclassical microwave drivers to produce an array of light beams with a multipartite entangled state. We have demonstrated that the resulting multipartite entanglement can be exploited to implement a quantum network. Furthermore, we have proven that in the presence of nonclassical driving microwave modes, teleportation between any pair of network nodes can be accomplished with a fidelity greater than the quantum threshold, enabling secure communication even in the presence of losses. The proposed scheme is a demonstration of a secure teleportation network implementing realistic lossy channels.\\

\section*{Appendix A}
The perturbed graphene conductivity expressions are  given by $\sigma_s^{(1)}=\frac{i e^2}{4\pi \hbar}{\rm ln\left\{\frac{4\pi\mu^{(1)}_c-\hbar(\omega+i 2\pi \Gamma )}{4\pi\mu^{(1)}_c+\hbar(\omega+i 2\pi \Gamma )}  \right\}}
+\frac {i 2 e^2 k_B T}{ \hbar^2(\omega+i 2\pi \Gamma )} \left\{\frac{\mu^{(1)}_c }{k_B T} +2{\rm ln\left\{1+e^{-\frac{\mu^{(1)}_c}{k_B T}}\right\}} \right\}$, and $\sigma_s^{(2)}=\frac{iq^2}{\pi\hbar}\frac{(\omega+i2\pi \Gamma)\hbar}{4(\mu^{(1)}_c)^2-(\omega+i2\pi \Gamma)^2\hbar^2}\mu^{(2)}_{c}+\frac{iq^2 K_B T}{\pi \hbar^2(\omega+j2\pi \Gamma} \\ \times tanh\bigg(\frac{\mu^{(1)}_{c}}{2K_B T}\bigg)  \frac{\mu^{(2)}_{c}}{K_B T}$ \cite{qsasexpress}. Here, $T$ is the operation temperature, $e$ is electron charge, $\Gamma=1/\tau$, and $\tau$ denotes the relaxation time. 

\section*{Appendix B}
\begin{figure}[h!]
\centering\includegraphics[width=0.8\columnwidth]{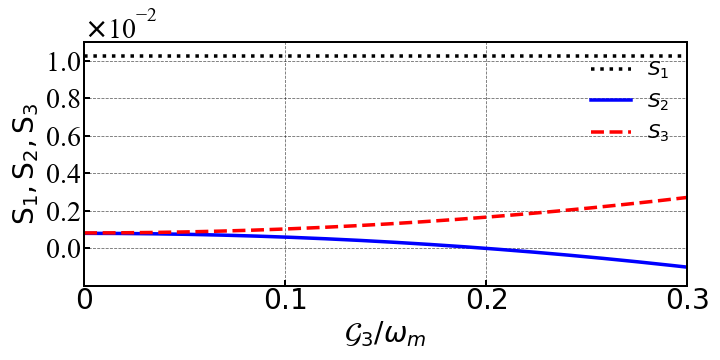} 
\caption{Stability conditions $S_1$ ,$S_2$ and $S_3$ as a function of the effective coupling $\mathcal{G}_3/\omega_m$ for all parameters same as in Fig.(\ref{fig2}). }
\label{fig5}
\end{figure}

Following the Routh-Hurwitz criterion, the steady-state stability conditions derived from  $\mathcal{A}^j$ in E.q. (12) are given by:
\begin{equation}
\begin{split}
&
S_1=\gamma^j_m+\gamma^j_2+\gamma^j_3>0, \\& S_2= \gamma^j_m+\mathcal{G}^{j^2}_2/ \gamma^j_2  -  \mathcal{G}^{j^2}_3/\gamma^j_3>0,
\\&S_3= \gamma^j_2+\gamma^j_3+\dfrac{ \mathcal{G}^{j^2}_2}{\gamma^j_m+\gamma^j_3} -   \dfrac{ \mathcal{G}^{j^2}_3}{\gamma^j_m+\gamma^j_2} >0.
\end{split}
\end{equation}
In Fig. (\ref{fig5}), we evaluate the stability parameters $S_1$, $S_2$ and $S_3$ against $\mathcal{G}_3/\omega_m$ while considering all other values same as in Fig.(\ref{fig2}). The simulations in Fig.(\ref{fig5}) show that the stability conditions are satisfied for $\mathcal{G}_3/\omega_m< 0.2$. Hence, the considered parameters in this work, including $\mathcal{G}_3/\omega_m= 0.14$, are corresponding to stable steady-state case.  

\section*{Appendix C}
\begin{figure}[h!]
\centering\includegraphics[width=0.8\columnwidth]{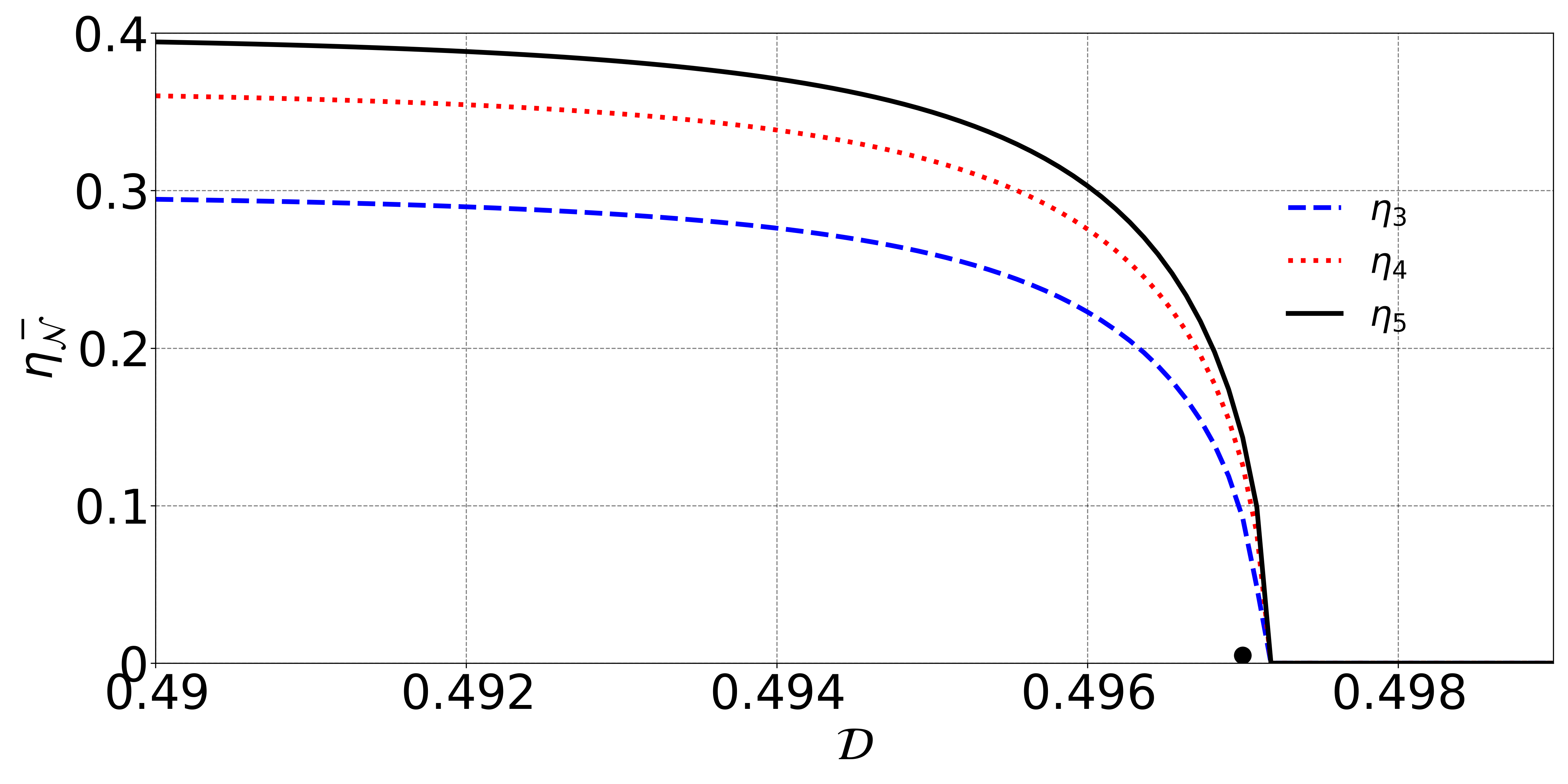} 
\caption{Plot of smallest symplectic eigenvalue $\eta^-_\mathcal{N}$ as a function of $\textit{nonclassicality depth}$ for all parameters same as in Fig.(\ref{fig2}).}
\label{fig6}

\end{figure}

The smallest symplectic eigenvalue $\eta^-_\mathcal{N}$ is calculated in Fig. (\ref{fig6}) as function of the nonclassicality depth $\mathcal{D}$. Different number of Bob's modes are considered. For instance, the blue dashed, red dotted, and black solid curves are for $\mathcal{N}=3,4$ and $5$, respectively. It can observed that the symplectic eigenvalue $\eta^-_\mathcal{N}$ approaches zero for $\mathcal{D}>0.4972$ (infinitely squeezed). It then follows that the logrithmic negativity $E_{\mathcal{N}}$ saturates for such  nonclassicality depth ranges (i.e., $\mathcal{D}>0.4972$). Therefore, in this work, we consider $\mathcal{D} \approx 0.497$ ( the black dot), which implies that $\eta^-_\mathcal{N}$ eigenvalue is just before approaching zero and no singularity is experienced.
$
\\
$

\noindent\textbf{Data Availability -} The data sets generated and/or analysed during the current study are available from the corresponding author on reasonable request.
$
\\
$

\noindent\textbf{Acknowledgments -} The authors thank Matteo G.A. Paris for the helpful discussions. This research is supported by Abu Dhabi Award for Research Excellence under ASPIRE/Advanced Technology Research Council (AARE19-062) 2019.
$
\\
$

\noindent\textbf{Competing Interests -} The authors declare that there are no competing interests. 
$
\\
$

\noindent\textbf{Correspondence -} Correspondence and requests for materials should be addressed to M.Q.~(email: montasir.qasymeh@adu.ac.ae).




\end{document}